\documentclass{article}

\PassOptionsToPackage{numbers, compress}{natbib}

\usepackage[preprint]{neurips_2023}

\usepackage[utf8]{inputenc} 
\usepackage[T1]{fontenc}    
\usepackage{hyperref}       
\usepackage{url}            
\usepackage{booktabs}       
\usepackage{amsfonts}       
\usepackage{nicefrac}       
\usepackage{microtype}      
\usepackage{xcolor}         

\hypersetup{
    colorlinks,
    linkcolor={red!50!black},
    citecolor={blue!50!black},
    urlcolor={blue!80!black}
}

\usepackage{amsmath}
\usepackage{graphicx}
\usepackage{caption}
\usepackage{algorithm}
\usepackage{algpseudocode}
\usepackage{multirow}
\usepackage{mathtools}
\usepackage{sidecap}
\usepackage{setspace}
\usepackage[caption=false]{subfig}

\sidecaptionvpos{figure}{m}

\newcommand{\bp}[1]{\vspace{0.2cm} \noindent \textbf{#1}:}

\title{Controllable Music Production with Diffusion Models and Guidance Gradients}

\author{%
  Mark Levy \\
  Apple \\
  \texttt{mark\_levy@apple.com} \\
  \And
  Bruno Di Giorgi \\
  Apple \\
  \texttt{bdigiorgi@apple.com} \\
  \And
  Floris Weers \\
  Apple \\
  \texttt{floris\_weers@apple.com} \\
  \And
  Angelos Katharopoulos \\
  Apple \\
  \texttt{a\_katharopoulos@apple.com} \\
  \And
  Tom Nickson \\
  Apple \\
  \texttt{tnickson@apple.com} \\
}

\begin{document}

\maketitle

\begin{abstract}
We demonstrate how conditional generation from diffusion models can be used to tackle a variety of realistic tasks in the production of music in 44.1kHz stereo audio with sampling-time guidance. The scenarios we consider include continuation, inpainting and regeneration of musical audio, the creation of smooth transitions between two different music tracks, and the transfer of desired stylistic characteristics to existing audio clips. We achieve this by 
applying guidance at sampling time in a simple framework that supports both reconstruction and classification losses, or any combination of the two. This approach ensures that generated audio can match its surrounding context, or conform to a class distribution or latent representation specified relative to any suitable pre-trained classifier or 
embedding model. Audio samples are available at \href{https://machinelearning.apple.com/research/controllable-music}{https://machinelearning.apple.com/research/controllable-music}.
\end{abstract}

\section{Introduction}

Recent work 
has shown great progress in addressing the challenging problem of generating musical audio with high enough quality for real world applications.
Language modelling approaches such as MusicLM and MusicGen \cite{https://doi.org/10.48550/arxiv.2301.11325, copet2023simple} tackle the problem of sequence length by working with compressed, tokenised representations originally developed for efficient audio encoding, and by cascading coarse and fine models to achieve realistic sounding audio at up to 32kHz.
Diffusion models such as Mo\^usai and Noise2Music \cite{2301.11757, https://doi.org/10.48550/arxiv.2302.03917} also show promising results,
again using a cascade of models.
These systems focus on conditional generation from 
descriptions of the desired content, such as ``\emph{a calming violin melody backed by a distorted guitar riff}''.
This
allows the creation of an impressive variety of sounds but
limits control over the musical output 
to generalised concepts, and conditioning relies on suitable paired data being available at training time.
In our work, we scale up waveform and latent diffusion to reach high audio quality,
and then explore some of the approaches to creative editing that become possible with a pre-trained diffusion model. 
Noting the similarity between the reconstruction guidance of \cite{chung2023diffusion} and the classifier guidance first introduced in \cite{dhariwal2021diffusion}, we combine both in a single framework, allowing us to tackle a wide range of useful music production tasks where control is provided in the form of example audio. 
Conditioning on audio prompts provides intuitive and fine-grained control over the musical characteristics of generated output, 
while applying conditioning with guidance gradients at sampling time removes the requirement to have paired data when training our diffusion models.
By analogy with previous similar work on controllable image modification, we see huge potential for music production incorporating a diffusion model as a generative prior, and our work only scratches the surface of possible methods and applications.

\section{Related work}

\subsection{Creative editing with diffusion models}
Diffusion models for high resolution images now provide the basis for a wide range of methods for creative editing of photos and videos. Inpainting can be implemented most simply by replacing the estimate of unmasked regions required during sampling with the original pixel values \cite{song2020generative}. This approach 
can be improved by fine-tuning with masking, or even training from scratch with a suitable masking strategy \cite{nichol2021glide, daras2023ambient}. Individual elements of an image can be edited starting from a rough guide, where for example part or all of the desired final image is provided in the form of a sketch or shape mask
\cite{meng2021sdedit}. If a paired dataset of sketch and image examples is available, then a separate control network can be fine-tuned to provide conditioning 
at sampling time \cite{zhang2023adding}. With models trained to generate images conditioned on the embedding of a text prompt, semantic editing can be implemented by modifying the prompt and generating pixels within an explicitly specified region, or by manipulating cross-attention maps between text and image
\cite{avrahami2022blended, kawar2023imagic, hertz2022prompt}.

A related line of research uses diffusion models to solve \emph{inverse problems}, where the aim is to reconstruct a complex signal as precisely as possible from a degraded version \cite{kawar2022denoising, chung2023diffusion, daras2023ambient}. While the motivation for these methods is reconstruction of missing or corrupted data, they can easily be applied to creative settings, where we wish to regenerate some part of an original signal.

\subsection{Diffusion models for musical audio}
Diffusion models have been applied to music by fine-tuning a pre-trained image model to generate audio spectrograms by treating them directly as images \cite{Rombach_2022_CVPR, Forsgren_Martiros_2022}. 
Most recent work focuses on diffusion models trained directly on audio signals to ensure higher quality results. CRASH works directly in the waveform domain to generate short drum hits at 44.1kHz \cite{rouard_crash_2021}. 
Noise2Music trains a cascade of diffusion models to generate 30 second clips of 24kHz audio, conditioned on text prompts learned from a large paired dataset of music and synthetic pseudo-labels 
\cite{https://doi.org/10.48550/arxiv.2302.03917}. Mo\^usai introduces a latent diffusion model for audio, using an independently trained diffusion autoencoder to compress spectrograms and then generating in the resulting latent domain \cite{preechakul2022diffusion, 2301.11757}. 
CQTDiff applies previous work on inverse problems to audio reconstruction with a diffusion model operating in the Constant-Q transform domain, including inpainting short sections of piano music recorded at 22kHz \cite{moliner_cqt-diff_2022}.

\section{Conditional generation with guidance gradients}
\label{section:conditional-generation}

Similarly to \cite{chung2022score}, 
we formalise conditional sampling as a multi-objective optimization problem:

\begin{equation}
\begin{aligned}
\max_{\mathbf{x}} \quad & J_1(\mathbf{x}) = p_{\text{data}}(\mathbf{x})\\
\min_{\mathbf{x}} \quad & J_2(\mathbf{x}) = d(\mathbf{y}, \mathcal{A}(\mathbf{x}))\\
\end{aligned}
\end{equation}

where $\mathbf{x} \in \mathbb{R}^n$ is the sample, $\mathcal{A} \in \mathbb{R}^n \rightarrow \mathbb{V}$ is a non-linear, differentiable measurement operator, $\mathbf{y} \in \mathbb{V}$ is the measured output and $d$ is a distance function $\mathbb{V} \times \mathbb{V} \rightarrow \mathbb{R}^+$. $J_1$ is maximised for samples that have high probability under the data distribution $p_{\text{data}}$ and is solved using the reverse process of a diffusion model. $J_2$ is minimized for samples that are consistent with the measurement $\mathbf{y}$.

Diffusion models solve the first sub-problem with an iterative algorithm, starting from noisy $\mathbf{x}_T \sim \mathcal{N}(\mathbf{0}, \mathbf{I})$ and refining the sample $\mathbf{x}_t$ at every iteration, with $t$ decreasing from $T$ to $0$. It is therefore a natural choice to solve the second problem by including a gradient descent step $\mathbf{x}_t \coloneqq \mathbf{x}_t - \xi \nabla_{\mathbf{x}_t} J_2(\mathbf{x}_t)$ at each iteration of the denoising process, where $\xi$ is the step size. This strategy can be interpreted as the alternation of unconditional updates and projections towards the measurement subspace \cite{chung2023diffusion}.

Depending on $\mathcal{A}$, specific variations can be applied. When $\mathcal{A}$ is defined for noiseless inputs only, a different ``denoising'' measurement operator $\mathcal{A}'$ can be used instead. $\mathcal{A}'$ is obtained as the composition of the original measurement operator $\mathcal{A}$ and the differentiable denoising operator of the diffusion model $\hat{\mathbf{x}}_0(\mathbf{x}_t)$, which estimates the noiseless sample at each iteration.
When $\mathcal{A}$ is linear $\mathcal{A}(\mathbf{x})=\mathbf{A}\mathbf{x}$ with $\mathbf{A} \in \mathbb{R}^{m \times n}$ having full row rank, we can apply a data consistency step $\mathbf{x}_t \coloneqq \mathbf{x}_t + \mathbf{A}^T (\mathbf{A}\mathbf{A}^T)^{-1} (\mathbf{y} - \mathbf{A}\mathbf{x}_t)$ at the end of each iteration, which exactly projects $\mathbf{x}_t$ onto the measurement subspace $\mathbf{y} = \mathbf{A}\mathbf{x}$. The resulting algorithms are provided in Section \ref{section:sampling_appendix} of the appendix.

\section{Experiments}

\subsection{Model architectures}

We train two classes of diffusion models, a waveform model and a latent
diffusion model. Our waveform model is a one-dimensional Unet
\cite{ronneberger2015u} with 440M parameters. 
Our latent diffusion model comprises a Variational Autoencoder (VAE) with a downsampling ratio of 128 in the time dimension and a
transformer diffusion model \cite{peebles2022dit} with 1B parameters. See Section \ref{section:architecture_appendix} of the appendix for full details.
As a baseline we also train a CQT diffusion model \cite{moliner_cqt-diff_2022}, using the authors' reference code.

\subsection{Dataset and metrics}

We train all our diffusion models on the Free Music Archive dataset \cite{defferrard2017fma},
which consists of 100k tracks totalling 8k hours of music. We hold out 80 hours as a test set.

To evaluate the quality of the generated audio we compute the Fr\'{e}chet Audio
Distance (FAD) \cite{kilgour2018fr} in the VGGish \cite{hershey2017cnn}
embedding space; and the KL Divergence \cite{iashin2021taming} in the AudioSet
class space \cite{7952261}, using the Patchout classifier \cite{Koutini_2022}.
Both metrics are computed with respect to reference statistics computed on the
training set.

\begin{table*}
\tiny
\resizebox{\textwidth}{!}{%
\begin{tabular}{lll|rr|rrr|rrr|r|rr|}
\toprule
{} & {} & {} & \multicolumn{2}{c|}{Unconditional} & \multicolumn{3}{c|}{Infill/Regeneration} & \multicolumn{3}{c|}{Continuation} & \multicolumn{1}{c|}{Transitions} & \multicolumn{2}{c|}{Cl. Guidance} \\
{Model} & {Sampling} & {Steps} & {FAD$\downarrow$} & {KLD$\downarrow$} & {FAD$\downarrow$} & {KLD$\downarrow$} & {MR$\downarrow$} & {FAD$\downarrow$} & {KLD$\downarrow$} & {MR$\downarrow$} & {Realism$\uparrow$} & {FAD$\downarrow$} & {KLD$\downarrow$} \\
\midrule

\multirow[t]{3}{*}{Latent} & DDIM & 50 
& 1.13 
& 0.33 
& 0.45/0.53 
& 0.19/0.19 
& 2.68/1.76
& 0.49 
& 0.22 
& 3.11$$
& 0.95$\pm$0.12
& 1.22 
& 0.30 \\

 & \multirow[t]{2}{*}{DDPM} & 50 
 & 1.10 
 & 0.34 
 & \bfseries 0.39/0.42 
 & \bfseries 0.17/0.17 
 & 2.63/2.30 
 & 0.50 
 & \bfseries 0.18 
 & 3.00$$
 & 0.95$\pm$0.12
 & 1.31 
 & 0.21 \\
 
 &  & 500 
 & \bfseries 0.72 
 & \bfseries 0.28 
 & 0.55/0.49 
 & 0.18/0.19 
 & 2.63/{\bfseries 1.72}
 & \bfseries 0.46 
 & 0.20 
 & 3.03$$
 & 0.94$\pm$0.10
 & \bfseries 0.96 
 & \bfseries 0.08 \\
 
\multirow[t]{2}{*}{Waveform} & \multirow[t]{2}{*}{DDPM} & 50 
& 8.42 
& 1.72 
& 1.36/0.47 
& 0.36/0.25 
& {\bfseries 1.67}/2.87 
& 2.73 
& 0.93 
& 4.42$$
 & 1.05$\pm$0.17
& 8.68 
& 1.57 \\

 &  & 500 
 & 4.35 
 & 1.35 
 & 0.56/0.58 
 & 0.28/0.27 
 & 2.84/2.69 
 & 0.77 
 & 0.72 
 & 3.82$$
 & 1.00$\pm$0.14
 & 6.28 
 & 0.30 \\
 
CQTDiff & DDPM & 50 
&  
&  
& 1.98/ -- \hspace{0.7em} 
& 0.42/ -- \hspace{0.7em} 
& 3.10/ -- \hspace{0.7em} 
&  
&  
&  
&
&  
&  \\

\midrule
VAE &  &  
& 0.47  
& 0.13  
& 0.51 
& 0.15 
& 1.48 
& 0.40 
& 0.12 
& 1.48
& 0.96$\pm$0.09
& 0.47 
& 0.13 \\

\midrule
Test Set &  &  
& 0.23  
& 0.04  
& 0.25 
& 0.08 
& 0.00 
& 0.18 
& 0.07 
& 0.00
& 1.00$\pm$0.00
& 0.23 
& 0.04 \\

\bottomrule
\end{tabular}
}
\caption{Experimental results} \label{tab:quant_results}

\end{table*}

\subsection{Creative applications}

We evaluate 
our models in a set of creative applications described below (further details in Section \ref{ssec:guidance-appendix} of the appendix). Results are shown in Table
\ref{tab:quant_results}.

\bp{Unconditional generation}
We sample
7k five second audio clips starting
from Gaussian noise. As a reference we include the scores obtained by
evaluating the clips in the test set and their reconstruction using the VAE, which serves as an upper bound for the Latent model.

\bp{Continuation}
We take the first 2.4s from each test set example, 
and use the model to reconstruct a possible continuation up to 6s. We include the mel-reconstruction distance \cite{garcia2023vampnet} (``MR''\ in Table \ref{tab:quant_results}) as a measure of consistency 
between the generated and the original continuation.

\bp{Infill / Regeneration}
An audio segment with duration equal to 6s is extracted
from each test set example. In the infill task, the middle 2 seconds of each segment are masked out and then generated using the model.
In the regeneration task the original audio is partially noised using the 
forward process, which ensures that basic rhythmic structure is maintained while other details are obscured, and used as the starting sample instead of Gaussian noise.
In both cases the left and right contexts are used to condition the generation
as done in the continuation task.

\bp{Transitions}
The transition task is a variant of the regeneration task: we start with two different 
tracks on the left and the right sides, and a 2.5s constant-power crossfade
between them in the middle segment to be regenerated.
Ideally we want the regenerated section to sound musical even when
the raw crossfade contains rhythmic and harmonic clashes.
We evaluate this with the realism score introduced in \cite{kynkaanniemi2019improved}, normalised track-by-track by the score of the constant-power crossfade.
We evaluate this task over a set of 100 transitions between randomly extracted pairs of tracks, with the reference manifold computed by projecting the training set on the Patchout classifier's embedding space \cite{Koutini_2022}.

\begin{SCfigure}
\includegraphics[width=0.55\textwidth, clip, trim=0cm 0 0cm 0]{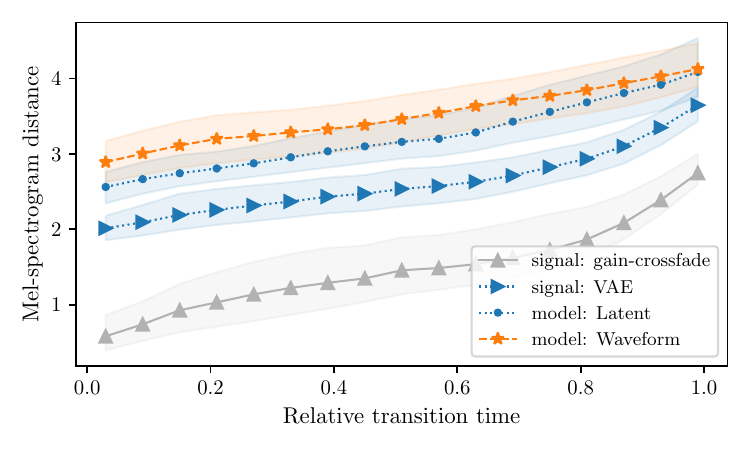}
\caption{Average mel-spectrogram distance of generated transitions with respect to the first track, over the duration of the transition. The distance increases linearly for all methods, implying smooth transitions. The ML-generated transitions start at non-zero distance because of the imperfect reconstruction.}
\centering
\label{fig:transition-mse}
\end{SCfigure}

We illustrate transition smoothness in Fig.\ \ref{fig:transition-mse}, which shows the average mel-reconstruction error relative to the first track over the duration of the transition.

\bp{Classifier guidance}
A common technique to improve the realism of conditionally generated samples
is to use pre-trained classifiers to compute the gradients.
We use the gradient of the L2 loss on the embedding space of the Patchout classifier \cite{Koutini_2022} to
guide our generation.

\subsection{Subjective evaluation}

We run a blind pairwise comparison test where we present the rater two samples generated by two models (or one sample vs the reference audio) for the same audio prompt and creative task. The raters are asked to choose the preferred sample from each pair based on perceived quality. 
We also encourage the reader to listen to the samples available at \href{https://machinelearning.apple.com/research/controllable-music}{https://machinelearning.apple.com/research/controllable-music}.

\begin{SCfigure}
\includegraphics[width=0.65\textwidth, clip, trim=0.6cm 0 1cm 0]{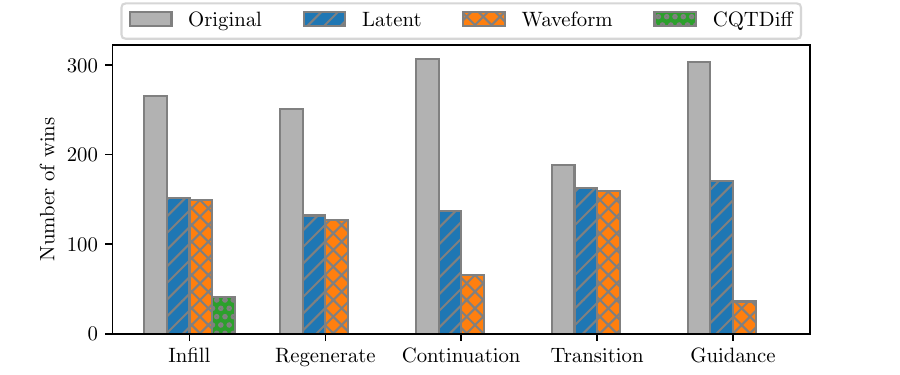}
\caption{Subjective evaluation results. We show the number of wins in head-to-head preference comparisons between samples generated by each class of model and also the original i.e. the prompt audio itself.}
\centering
\label{fig:wins}
\end{SCfigure}

\subsection{Discussion}
Results in Table \ref{tab:quant_results} show that, regardless
of the sampling method, the model
operating in a latent space generally produces musical audio that is 
higher quality
and closer to the 
original
than 
the waveform diffusion model. 
This general trend confirms that cascading generation
is beneficial, as previously suggested in \cite{https://doi.org/10.48550/arxiv.2302.03917}.
Both models improve over infilling with the baseline CQTDiff model by a large margin.
We summarise the results of the subject evaluation in Fig. \ref{fig:wins}. These confirm the superiority of our models over the baseline
and suggest that the latent representation gives an advantage in tasks where a data consistency step is unavailable (guidance) or weaker (continuation,
where we have only a one-sided prompt).

\section{Conclusions}

In this paper we explored a simple framework that enables different applications of diffusion generative models in the context of
high-fidelity music production. Applications such as continuation, inpainting, regeneration, transition and latent conditioning have been evaluated for two model architectures showing the relative importance of architectural and sampling choices for the different tasks.

\vfill\pagebreak

\bibliographystyle{unsrt}
\bibliography{generation}

\vfill\pagebreak

\appendix

\section{Other related work}

\subsection{Language models for music generation}
We focus in this paper on diffusion models, but there is a growing body of parallel work on language modelling for musical audio, operating in the discrete token space of VQ-VAEs. Jukebox uses a cascade of three transformers trained with metadata and song lyrics as well as encoded audio, to enable generation conditioned on song texts and artist identity \cite{dhariwal_jukebox_2020}. PerceiverAR and MusicGen model interleaved sequences of audio tokens from the multiple residual VQ-VAE codebooks of neural codecs, while MusicLM extends this approach by conditioning on ``semantic" tokens from a pre-trained first stage model which models audio at a lower sample rate \cite{hawthorne2022generalpurpose, https://doi.org/10.48550/arxiv.2301.11325, copet2023simple, chung2021w2vbert}. MusicLM is conditioned on text captions and also on melody, by concatenating automatically extracted melody tokens with the text tokens, and this supports generation of samples conditioned on a provided melody as well as a text prompt. 

In contrast to these autoregressive models, VampNet introduces a language model trained on a masked token prediction objective which supports inpainting, and also employs an efficient iterative sampling scheme to generate interleaved codebook tokens \cite{garcia2023vampnet, chang2022maskgit}. The authors experiment with enforcing the rhythmic consistency of generated sections by retaining tokens around the expected beat positions in the region to be inpainted.

\section{Conditional generation} \label{section:sampling_appendix}

\subsection{Diffusion models}

A Gaussian denoising diffusion probabilistic model (DDPM) defines a forward corruption process through time, which gradually applies noise to real data $\mathbf{x}_0 \sim p_\text{data}$
\begin{equation*}
q(\mathbf{x}_t | \mathbf{x}_0) = \mathcal{N}(\mathbf{x}_t; \alpha_t\mathbf{x}_0, \sigma_t^2\mathbf{I}),
\end{equation*}
where $\alpha_t$ and $\sigma_t$ are constants defining a noise schedule with monotonically decreasing signal-to-noise ratio $\alpha_t / \sigma_t$,  such that $\mathbf{x}_T \sim \mathcal{N}(\mathbf{0}, \mathbf{I})$. We use a cosine schedule with continuous $t \in \left[ 0, 1 \right]$, $\alpha_t = \cos(\pi t / 2)$ and $\alpha_t^2 + \sigma_t^2 = 1$. 
We can sample noised data directly with 
$\mathbf{x}_t = \alpha_t \mathbf{x}_0 +  \sigma_t \boldsymbol{\epsilon}_t$
and $\boldsymbol{\epsilon}_t \sim \mathcal{N}(\mathbf{0}, \mathbf{I})$.

DDPMs are trained to learn a reverse process 
$p_\theta(\mathbf{x}_s | \mathbf{x}_t) = \mathcal{N}(\mu_\theta(\mathbf{x}_t, t), \sigma_{t \rightarrow s})$ where 
$\sigma_{t \rightarrow s} = (\sigma^2_s / \sigma^2_t)(1 - \alpha^2_t / \alpha^2_s)$ are constants computed from the noise schedule and $\mu_\theta$ is the output of a denoising model that estimates $\mathbf{x}_0$ or equivalently $\boldsymbol{\epsilon}_t$. We choose the so called \emph{v prediction} parameterization for our estimator, learning to predict a combination of the added noise and the original signal
$\mathbf{v}_t = \alpha_t \boldsymbol{\epsilon}_t - \sigma_t \mathbf{x}_0$ \cite{salimans2021progressive}.
We train our model to minimise a reconstruction loss
\begin{equation*}
L(\theta) = \mathbb{E}_{t \sim \mathcal{U}(0, 1), \mathbf{x}_0 \sim p_\text{data}, \boldsymbol{\epsilon} \sim \mathcal{N}(0, \mathbf{I})} \left[ \big\| \mathbf{v}_t - v_{\theta}(\mathbf{x}_t, t) \big\|^2 \right].
\end{equation*}

Sampling from a DDPM starts with Gaussian noise $\mathbf{x}_T$ and uses $p_\theta(\mathbf{x}_s|\mathbf{x}_t)$ to produce gradually less noisy samples until reaching a final sample $\mathbf{x}_0$.
An important result from \cite{song2020score} is that the output of the denoising model is equivalent to estimating an evidence lower bound on the \emph{score function}
$\nabla_{\mathbf{x}_t} \log{p(\mathbf{x}_t})$, which enables the use of samplers derived from denoising score matching, such as DDIM \cite{song2020denoising}.

\subsection{Sampling with guidance gradients}
\label{ssec:guidance-appendix}

With reference to Sect.\ \ref{section:conditional-generation}, Algorithms \ref{alg:guidance} and \ref{alg:guidance_DDIM} illustrate the DDPM and DDIM iterative methods for conditional generation. 
We model different conditional generation tasks by varying specific parameters of the 
algorithm: the measurement operator $\mathcal{A}$, the distance function $d$ and the starting sample $\mathbf{x}_T$, as shown in Table \ref{tab:task-parameters}.

\begin{table*}[ht]
    \tiny
    \centering
    \begin{tabular}{|l|l|l|l|l|l|l|}
         \toprule
         Task & $\mathcal{A}(\mathbf{x})$ & $d(\mathbf{y}, \mathcal{A}(\mathbf{x}))$ & $\mathbf{y},\quad \mathbf{x}_T$ \\
         \midrule
         \midrule
         Continuation & 
         $ {\mathcal{A}(\mathbf{x}) = \mathbf{A}\mathbf{x}, \quad  \mathbf{A} = \mathbf{A_L}} $ & $ \| \mathbf{y} - \mathbf{A}\mathbf{x}\|_1 $ & 
         $ \begin{aligned} & \mathbf{y} = \mathbf{A}\mathbf{\bar{x}} \\ & \mathbf{x}_T = \mathbf{A}^T\mathbf{y} + (\mathbf{I} - \mathbf{A}^T\mathbf{A})\mathbf{z} \end{aligned}$ \\
         \midrule
         Infill & 
         $ {\mathcal{A}(\mathbf{x}) = \mathbf{A}\mathbf{x}, \quad  \mathbf{A} = \left[{\begin{aligned}& \mathbf{A_L}\\ & \mathbf{A_R} \end{aligned}}\right]} $ & 
         as Continuation & 
         as Continuation \\
         \midrule
         Regenerate & 
         as Infill & 
         as Continuation & 
         $\begin{aligned} & \mathbf{y} = \mathbf{A}\mathbf{\bar{x}} \\ & \mathbf{x}_T = \mathbf{A}^T\mathbf{y} + (\mathbf{I} - \mathbf{A}^T\mathbf{A})(k\mathbf{z} + (1-k)\mathbf{\bar{x}}) \end{aligned}$ 
         \\
         \midrule
         Transition & 
         as Infill & 
         as Continuation & 
         $\begin{aligned} 
         & \mathbf{\bar{x}} = \left[{\begin{aligned}& \mathbf{A_L}\mathbf{\bar{x}}_L\\ & \mathbf{F}_\text{out}\mathbf{\bar{x}}_L +  \mathbf{F}_\text{in}\mathbf{\bar{x}}_R \\ & \mathbf{A_R}\mathbf{\bar{x}}_R \end{aligned}}\right] \\ 
         & \mathbf{y} = \mathbf{A}\mathbf{\bar{x}} \\
         & \mathbf{x}_T = \mathbf{A}^T\mathbf{y} + (\mathbf{I} - \mathbf{A}^T\mathbf{A})(k\mathbf{z} + (1-k)\mathbf{\bar{x}}) \\
         & \mathbf{F}_\text{out} = [\mathbf{0}, \text{diag}(\mathbf{f}_\text{out}), \mathbf{0}] \\
         & \mathbf{F}_\text{in} = [\mathbf{0}, \text{diag}(\mathbf{f}_\text{in}), \mathbf{0}] \\         
         \end{aligned}$
         \\
         \midrule
         Embedder guidance & $\mathcal{A} \in \mathrm{R}^n \rightarrow \mathrm{R}^m$ & $\| \mathbf{y} - \mathbf{A}\mathbf{x}\|_2$ & $
         \mathbf{y} = \mathcal{A}(\mathbf{\bar{x}}),\quad \mathbf{x}_T = \mathbf{z}$ \\
         \midrule
         Classifier guidance & $\mathcal{A}(\mathbf{x})_i = p(c_i | \mathbf{x}) \in [0, 1]^m$ & $\text{BCE}(\mathbf{y}, \mathcal{A}(\mathbf{x}))$ & $ \mathbf{y} \in [0, 1]^m,\quad \mathbf{x}_T = \mathbf{z}$ \\
         \bottomrule
    \end{tabular}
    \caption{Task specific parameters. $\mathbf{A}_L = [\mathbf{I}_{C_L}, \mathbf{0}] \in \mathrm{R}^{C_L \times n}$, $\mathbf{A}_R = [\mathbf{0}, \mathbf{I}_{C_R}] \in \mathrm{R}^{C_R \times n}$; $C_L$ and $C_R$ are the sample lengths of the left and right contexts respectively, $C_L + C_R < n$; $\mathbf{\bar{x}}$, $\mathbf{\bar{x}}_L$, $\mathbf{\bar{x}}_R$ are target signals;  $\mathbf{z}$ is a noise signal $\mathbf{z} \sim \mathcal{N}(\mathbf{0}, \mathbf{I})$, $k \in [0, 1]$ is a scalar coefficient that regulates the amount of noise in the infill region of the initial sample for the regenerate and transition tasks, $\text{BCE}$ stands for Binary Cross-Entropy. In the transition task $\mathbf{f}_\text{out}$ and $\mathbf{f}_\text{out}$ represent fade out and fade in coefficients of a constant-power cross-fade. The audio channel dimension is not considered for simplicity.}
    \label{tab:task-parameters}
\end{table*}

\algnewcommand{\LineComment}[1]{\State \(\triangleright\) #1}

\begin{algorithm}[H]
\caption{Sampling with guidance gradients}\label{alg:guidance}
\begin{algorithmic}
\Require initial sample $\mathbf{x}_T$, measurement $\mathbf{y}$, distance function $d$, measurement operator $\mathcal{A}(\cdot)$, gradient step size $\xi$, $N$ timesteps $t_i$ in $(T, 0)$

\For {$i =N$ \textbf{to} $1$}
	\State $t \gets t_i$, $s \gets t_{i-1}$

    \vspace{0.3em}
	\State $\hat{\mathbf{x}}_0 \gets \alpha_t \mathbf{x}_t - \sigma_t v_\theta(\mathbf{x}_t, t)$
	
    \vspace{0.4em}
	\State $\boldsymbol{\epsilon} \sim \mathcal{N}(0, \mathbf{I})$
 
	\State $\mathbf{x}_s \gets \cfrac{\alpha_s}{\sigma^2_t} (1 - \cfrac{\alpha^2_t}{\alpha^2_s}) \hat{\mathbf{x}}_0
	+
	\cfrac{\alpha_t \sigma^2_s}{\alpha_s \sigma^2_t}\mathbf{x}_t
	+
	\sigma_{t \rightarrow s} \boldsymbol{\epsilon}$

    \vspace{0.3em}
	\State $\mathbf{x}_s \gets \mathbf{x}_s - \xi \nabla_{\mathbf{x}_t} d(\mathbf{y}, \mathcal{A}(\mathbf{x}_t))$ 
    \Comment{Alternatively $\mathcal{A}(\mathbf{\hat{x}}_0)$}
    
    \vspace{0.4em}
    \LineComment{Apply data consistency step if possible (see Sect.\ \ref{section:conditional-generation})}
\EndFor

\State \Return $\mathbf{x}_0$

\end{algorithmic}
\end{algorithm}

\begin{algorithm}[H]
\caption{DDIM with guidance gradients}\label{alg:guidance_DDIM}
\begin{algorithmic}
\Require initial sample $\mathbf{x}_T$, measurement $\mathbf{y}$, distance function $d$, measurement operator $\mathcal{A}(\cdot)$, gradient step size $\xi$, $N$ timesteps $t_i$ in $(T, 0)$

\For {$i =N$ \textbf{to} $1$}
    \vspace{0.3em}
	\State $t \gets t_i$, $s \gets t_{i-1}$

    \vspace{0.3em}
	\State $\hat{\mathbf{x}}_0 \gets \alpha_t \mathbf{x}_t - \sigma_t v_\theta(\mathbf{x}_t, t)$

    \vspace{0.3em}
    \State $\hat{\boldsymbol{\epsilon}}_t \gets (\mathbf{x}_t - \alpha_t \hat{\mathbf{x}}_0) / \sigma_t$

    \vspace{0.3em}
    \State $\hat{\boldsymbol{\epsilon}}_t \gets \hat{\boldsymbol{\epsilon}}_t - \xi \sigma_t \nabla_{\mathbf{x}_t} d(\mathbf{y} - \mathcal{A}(\mathbf{x}_t))$
    \Comment{Alternatively $\mathcal{A}(\mathbf{\hat{x}}_0)$}

    \vspace{0.3em}
    \State $\mathbf{x}_s \gets \alpha_s \hat{\mathbf{x}}_0 + \sqrt{(1 - \alpha^2_s)} \hat{\boldsymbol{\epsilon}}_t$
    
    \vspace{0.4em}
    \LineComment{Apply data consistency step if possible (see Sect.\ \ref{section:conditional-generation})}
\EndFor

\State \Return $\mathbf{x}_0$

\end{algorithmic}
\end{algorithm}

\section{Model architecture and training details} \label{section:architecture_appendix}

\subsection{Models}
Our waveform diffusion model uses a similar Unet
implementation to \cite{rombach_stablediffusion_2022} to generate 11.9 seconds of 44.1kHz stereo audio.
To deal efficiently with the large input size we ``fold'' the 2-dimensional stereo
input into the channel dimension and ``unfold'' the output. We minimize possible
aliasing artifacts by using overlapping windows with frame size 32 and hop size 16, and multiplying with a hamming
window function before aggregating.

Our latent diffusion model comprises a Variational Autoencoder (VAE) to extract
the latent representations and a transformer diffusion model to generate 5.9 seconds of audio. Contrary to
previous works on images, we use a much higher downsampling ratio for our VAE
to make the input size tractable for the transformer. In particular, our
VAE consists of 8 layers in the encoder and decoder. After every layer but the
first one we use a $2 \times$ down- or upsampling operation respectively,
resulting in a latent sequence $128 \times$ smaller than the input. The layers
are implemented as 1D ConvNext \cite{liu2022convnet} layers, which we observed
perform better than the traditional dilated convolutions. Finally, we train our
VAE using frequency losses as well as L1 and L2 losses
\cite{yamamoto2020parallel}. To improve the fidelity of the audio reconstructed
from the latent space, we finetune the decoder while keeping the encoder fixed
using a multi-scale frequency based discriminator for a small number of steps (similar to \cite{defossez_high_2022}).
Our transformer model for latent diffusion follows \cite{peebles2022dit} closely,
the main difference being the addition of a relative
positional encoding \cite{shaw2018self}. We use 32 layers with 1536 feature
dimensions resulting in 1B parameters.

\subsection{Training}
The models are trained with the AdamW optimizer, $\beta_1 = 0.9$, $\beta_2 = 0.999$, and no weight decay. A cosine learning rate schedule with a warmup of 5000 steps is used at the beginning of training. Training uses fp16 mixed precision and distributed data parallelism. The waveform model is trained for 600k steps with a batch size of 384 on 24 A100 GPUs, and the latent model for 300k steps with a batch size of $96$ on 8 A100 GPUs.

\subsection{Sampling parameters}
We set the guidance gradient step size $\xi$ differently for latent ($\xi=3 \times 10^{-2}$) and waveform ($\xi=3 \times 10^{-3}$) models, based on informal qualitative evaluation. The waveform models appeared more sensitive to $\xi$ and degraded to generating noisy signals for higher values $\xi > 3 \times 10^{-2}$. We kept the step size fixed throughout the sampling process. We created initial samples for the regenerate and transition tasks using a noise mixing coefficient $k=0.85$.

\end{document}